\documentstyle[aps,preprint,12pt]{revtex}

\preprint{EFUAZ FT-96-34}

\begin{document}

\title{A Note on the Relativistic Covariance of the ${\bf B}-$
Cyclic Relations\thanks{Submitted to ``Foundation of
Physics Letters"}}

\author{{\bf Valeri V. Dvoeglazov}}

\address{
Escuela de F\'{\i}sica, Universidad Aut\'onoma de Zacatecas \\
Antonio Doval\'{\i} Jaime\, s/n, Zacatecas 98068, ZAC., M\'exico\\
Internet address:  VALERI@CANTERA.REDUAZ.MX}

\date{September 18, 1996. Revised May 7, 1997}

\maketitle

\begin{abstract}
It is shown that the Evans-Vigier modified electrodynamics
is compatible with the Relativity Theory.

\end{abstract}


\baselineskip32pt


Recently a new version of non-Maxwellian theories of
electromagnetism has been proposed~\cite{EV1,EV2}. As a matter of fact, the
Evans-Vigier ${\bf B}^{(3)}$ theory includes a spin variable in the
classical theory and presents itself straightforward development of the
Belinfante, Ohanian and Kim ideas~\cite{Belin,Ohan,Kim}. In the present
note I restrict myself only one particular question of the relativistic
covariance of this theory. I would not like to speak here about a numerous
variety of other generalizations of the Maxwell's theory referring a
reader to the recent review~\cite{DVOE1}. All those theories are earlier
given either  strong critics (while not always perfectly reasonable) or
ignorance and only in the nineties several new versions appeared at once,
what ensures that the question would obtain serious, careful and justified
consideration. The ${\bf B}^{(3)}$ model is not an exception. A list of
works criticizing this theory was presented in ref.~\cite{Comay2} and that
author wrote several critical comments too~\cite{Comay1,Comay2}. A serious
objection to the Evans-Vigier theory which was presented by Comay is that
he believes that the modified electrodynamics is not a relativistic
covariant theory. Questions of Dr. Comay may arise in future analyses of
the ${\bf B}^{(3)}$ theory  because he correctly indicated
some notational misunderstandings in the Evans and Vigier works.
Therefore, they are required detailed answers.

According to~\cite[Eq.(11.149)]{Jackson} the Lorentz transformation rules
for electric and magnetic fields are the following:
\begin{mathletters}
\begin{eqnarray}\label{l1}
{\bf E}^\prime &=& \gamma ({\bf E} + c {\bbox \beta} \times {\bf B} )
-{\gamma^2 \over \gamma+1} {\bbox\beta} ({\bbox\beta}\cdot {\bf E})\quad,\\
\label{l2}
{\bf B}^\prime &=& \gamma ({\bf B} -{\bbox \beta} \times {\bf E}/c )
-{\gamma^2 \over \gamma+1} {\bbox\beta} ({\bbox\beta}\cdot {\bf B})\quad
\end{eqnarray}
\end{mathletters}
where ${\bbox\beta} ={\bf v}/c$\,,\,\,$\beta = \vert {\bbox\beta}\vert
= \mbox{tanh} \phi$\,,\, $\gamma ={1\over \sqrt{1-{\bbox\beta}^2}}
=\mbox{cosh} \phi$, with $\phi$ being the parameter of the Lorentz boost.
We shall further use the natural unit system $c=\hbar =1$. After
introducing the spin matrices $({\bf S}_i)_{jk} =-i\epsilon_{ijk}$ and
deriving
relevant relations:
$$({\bf S}\cdot {\bbox \beta})_{jk} {\bf a}_k \equiv i
[{\bbox \beta} \times {\bf a}]_j\quad,$$ $${\bbox \beta}_j {\bbox \beta}_k
\equiv [ {\bbox \beta}^{\,2} - ({\bf S}\cdot {\bbox \beta})^2
]_{jk}\quad$$
one can rewrite Eqs. (\ref{l1},\ref{l2}) to the form
\begin{mathletters} \begin{eqnarray}\label{l3} {\bf E}^\prime_i &=& \left
( \gamma \openone + {\gamma^2 \over \gamma +1} \left [ ({\bf S}\cdot {\bbox
\beta})^2 -{\bbox\beta}^{\,2} \right ] \right )_{ij} {\bf E}_j -i\gamma
({\bf S}\cdot {\bbox\beta})_{ij} {\bf B}_j\quad,\\ \label{l4} {\bf
B}^\prime_i &=& \left ( \gamma \openone + {\gamma^2 \over \gamma +1} \left
[ ({\bf S}\cdot {\bbox \beta})^2 -{\bbox\beta}^{\,2} \right ] \right
)_{ij} {\bf B}_j +i\gamma ({\bf S}\cdot {\bbox\beta})_{ij} {\bf E}_j\quad.
\end{eqnarray}
\end{mathletters}
First of all, one should mention that these equations  are valid for
electromagnetic fields of various polarization configurations.
Next, Eqs. (\ref{l3},\ref{l4}) preserve
properties of the vectors ${\bf B}$ (axial) and ${\bf E}$ (polar) with
respect to the space inversion operation.  Furthermore, if we consider
other field configurations like $\phi_{_{L,R}} = {\bf E}\pm i{\bf B}$ or
${\bf B}\mp i{\bf E}$, the Helmoltz bivectors, which may already not have
definite properties with respect to the space inversion operation (namely,
they transform as $\phi_{_{R}} \leftrightarrow \pm \phi_{_{L}}$) we
obtain
\begin{equation} ({\bf B}^\prime \pm i{\bf E}^\prime)_i = \left ( 1
\pm \gamma ({\bf S}\cdot {\bbox \beta}) +{\gamma^2 \over \gamma +1} ({\bf
S}\cdot {\bbox \beta})^2 \right )_{ij} ({\bf B} \pm i{\bf E})_j\quad,
\end{equation}
i.e, it becomes obviously that
they transform as the right-  and the left- parts
of the Weinberg's $2(2S+1)-$ component field function~\cite{Weinberg}.

Now, we can consider the question of the Lorentz transformations for
transversal modes ${\bf B}^{(1)}$, ${\bf B}^{(2)}$, ${\bf E}^{(1)}$
and ${\bf E}^{(2)}$ and, hence, make a correct conclusion about
the transformation of ${\bf B}^{(1)}\times {\bf B}^{(2)} = {\bf E}^{(1)}
\times {\bf E}^{(2)} =
i B^{(0)} {\bf B}^{(3)\,\ast}$. In the first frame transversal modes of
the electromagnetic field have the following explicit forms,
ref.~\cite{EV3}, $\phi =\omega t -{\bf k}\cdot {\bf r}$:
\begin{mathletters}
\begin{eqnarray} \label{t1}
{\bf B}^{(1)} &=& {B^{(0)} \over \sqrt{2}} (i{\bf i} + {\bf j})
e^{i\phi}\quad,\quad
{\bf E}^{(1)} = -i {\bf B}^{(1)} = {E^{(0)} \over \sqrt{2}} ({\bf i} - i
{\bf j}) e^{i\phi}\quad,\quad \\
\label{t2}
{\bf B}^{(2)} &=& {B^{(0)} \over
\sqrt{2}} (- i{\bf i} + {\bf j}) e^{-i\phi}\quad,\quad
{\bf E}^{(2)} = +i {\bf B}^{(2)} =
{E^{(0)} \over \sqrt{2}} ({\bf i} + i {\bf j}) e^{-i\phi}\quad.
\end{eqnarray} \end{mathletters}
We have implied that in a free-space circularly-polarized radiation
$B^{(0)} = E^{(0)}$. The pure Lorentz  transformations (without inversions)
do not change  the sign of the phase of the field functions, so we should
consider separately properties of the set of ${\bf B}^{(1)}$ and ${\bf
E}^{(1)}$, which can be regarded as
the negative-energy solutions in QFT (cf. the Dirac case~\cite{Itzyk}),
and another set of ${\bf B}^{(2)}$ and ${\bf E}^{(2)}$,
as the positive-energy solutions. The opposite interpretation
is also possible and, in fact, was used by E. Comay ($\phi \rightarrow
- \phi$).  But these issues are indicated not to be relevant to
the present discussion. Thus, in this framework one
can deduce from Eqs. (\ref{l3},\ref{l4})
\begin{mathletters}
\begin{eqnarray} \label{l11} {\bf B}^{(1)\,\prime}_i &=& \left ( 1+
{\gamma^2 \over \gamma +1} ({\bf S}\cdot {\bbox \beta})^2 \right )_{ij}
{\bf B}^{(1)}_j +i\gamma ({\bf S}\cdot {\bbox\beta})_{ij} {\bf E}^{(1)}_j
\quad,\\
\label{l21}
{\bf B}^{(2)\,\prime}_i &=& \left ( 1
+{\gamma^2 \over \gamma +1} ({\bf S}\cdot {\bbox \beta})^2 \right )_{ij}
{\bf B}^{(2)}_j +i\gamma ({\bf S}\cdot {\bbox\beta})_{ij} {\bf E}^{(2)}_j
\quad,\\
\label{l31}
{\bf E}^{(1)\,\prime}_i &=& \left ( 1+
{\gamma^2 \over \gamma +1} ({\bf S}\cdot {\bbox \beta})^2 \right )_{ij}
{\bf E}^{(1)}_j -i\gamma ({\bf S}\cdot {\bbox\beta})_{ij} {\bf B}^{(1)}_j
\quad,\\
\label{l41}
{\bf E}^{(2)\,\prime}_i &=& \left ( 1
+{\gamma^2 \over \gamma +1} ({\bf S}\cdot {\bbox \beta})^2 \right )_{ij}
{\bf E}^{(2)}_j -i\gamma ({\bf S}\cdot {\bbox\beta})_{ij} {\bf B}^{(2)}_j
\quad,
\end{eqnarray} \end{mathletters}
Using relations between transversal modes of electric and magnetic
field (\ref{t1},\ref{t2}) one can formally write\footnote{One would
wish to study properties of this physical system with respect to the
space inversion operation. Since, explicit forms of transversal modes of
electric field in the first frame are proportional (with imaginary
coefficients) to the transversal modes of the magnetic
field (\ref{t1},\ref{t2}) some
fraction of ${\bf E}^{(k)}$ or ${\bf B}^{(k)}$ can be
formally substituted by the
vector of other parity (like we are doing in the process of calculations).
Furthermore, one can take any combinations of Eq.  (\ref{l11}) and Eq.
(\ref{l31}) multiplied by an arbitrary phase factor, or that of Eq.
(\ref{l21}) and Eq.  (\ref{l41}) multiplied by a phase factor. The parity
properties of the field functions in the general case would be different
in the left-hand side and in the right-hand side of resulting equations.
Generally speaking, the notation (\ref{l111}-\ref{l411})
is used in this paper only for simplification of calculations.
In fact, one can also proceed further with the forms
(\ref{l11}-\ref{l41}).}
\begin{mathletters} \begin{eqnarray} \label{l111}
{\bf B}^{(1)\,\prime}_i &=& \left ( 1+\gamma ({\bf S}\cdot {\bbox\beta})
+{\gamma^2 \over \gamma +1} ({\bf S}\cdot {\bbox \beta})^2 \right )_{ij}
{\bf B}^{(1)}_j \quad,\\ \label{l211} {\bf B}^{(2)\,\prime}_i &=& \left (
1 -\gamma ({\bf S}\cdot {\bbox\beta}) +{\gamma^2 \over \gamma +1} ({\bf
S}\cdot {\bbox \beta})^2 \right )_{ij} {\bf B}^{(2)}_j \quad,\\
\label{l311} {\bf E}^{(1)\,\prime}_i &=& \left ( 1+ \gamma ({\bf S}\cdot
{\bbox\beta}) +{\gamma^2 \over \gamma +1} ({\bf S}\cdot {\bbox \beta})^2
\right )_{ij} {\bf E}^{(1)}_j \quad,\\ \label{l411} {\bf
E}^{(2)\,\prime}_i &=& \left ( 1- \gamma ({\bf S}\cdot {\bbox \beta})
+{\gamma^2 \over \gamma +1} ({\bf S}\cdot {\bbox \beta})^2 \right )_{ij}
{\bf E}^{(2)}_j \quad,
\end{eqnarray} \end{mathletters}
We still observe that ${\bf B}^{(2)}$ can be related with
${\bf B}^{(1)}$ by the unitary matrix:
\begin{equation}
{\bf B}^{(2)} = U {\bf B}^{(1)} = e^{-2i\phi} \pmatrix{0&-i&0\cr
-i&0&0\cr 0&0&1\cr} {\bf B}^{(1)}\quad.
\end{equation}
Since this unitary transformation results in the change of the
basis of spin operators only, we deduce that the concepts of
properties of some geometrical object with respect to
Lorentz transformations and with respect to space-inversion
transformations can be simultaneously
well-defined concepts only after defining corresponding ``bispinors"
of the $(j,0)\oplus (0,j)$ representations
and keeping the same spin basis for both parts of the bispinor. See
also~\cite{Ryder} for the example in the $(1/2,0)\oplus (0,1/2)$ rep.

We still advocate that a) the  properties ${\bf E}^{(k)\,\prime}$
to be proportional to ${\bf B}^{(k)\,\prime}$ with
imaginary coefficients are preserved and b) ${\bf B}^{(1)}$ and ${\bf
E}^{(1)}$ in Eqs. (\ref{l111}-\ref{l411}) transform like ${\bf B} +i {\bf
E}$ of the Cartesian basis,  i.e., like the right part of the Weinberg
field function and ${\bf B}^{(2)}$ and ${\bf E}^{(2)}$, like ${\bf B} -
i{\bf E}$,  i.e., like the left part of the Weinberg field function.
Using the above rules to find the transformed 3-vector ${\bf
B}^{(3)\,\prime}$ is just an algebraic exercise.  Here it is
\begin{equation}
{\bf B}^{(1)\,\prime} \times {\bf B}^{(2)\,\prime} =
{\bf E}^{(1)\,\prime} \times {\bf E}^{(2)\,\prime} =  i\gamma (B^{(0)})^2
(1- {\bbox \beta} \cdot \hat{\bf k}) \left [ \hat {\bf k} -\gamma {\bbox
\beta} +{\gamma^2  ({\bbox \beta}\cdot \hat {\bf k}) {\bbox \beta}\over
\gamma+1} \right ]\quad,\label{ltt}
\end{equation}
where $\hat {\bf k}$ is
the orth vector of the axis $OZ$. We know  that the longitudinal mode in
the Evans-Vigier theory is defined as ${\bf B}^{(3)} =
{\bf B}^{(3)\,\ast} = B^{(0)} {\bf k}$.
Thus, considering that $B^{(0)}$ transforms as zero-component of the
four-vector and ${\bf B}^{(3)}$ as space components of the
four-vector:~\cite[Eq.(11.19)]{Jackson}
\begin{mathletters}
\begin{eqnarray}\label{lt1}
B^{(0)\,\prime} &=& \gamma (B^{(0)} -{\bbox\beta}\cdot {\bf B}^{(3)}
)\quad,\\
\label{lt2} {\bf B}^{(3)\,\prime} &=& {\bf B}^{(3)} + {\gamma -1 \over
\beta^2} ({\bbox \beta} \cdot {\bf B}^{(3)}) {\bbox\beta} - \gamma
{\bbox\beta} B^{(0)} \quad, \end{eqnarray} \end{mathletters}
we find  from
(\ref{ltt}) that the relation between transversal and longitudinal modes
preserves its form:  \begin{equation} {\bf B}^{(1)\,\prime} \times {\bf
B}^{(2)\,\prime} = i B^{(0)\,\prime} {\bf B}^{(3)\, \ast \, \prime}\quad.
\end{equation}
A reader interested in these matters can exercise to prove the covariance
of other cyclic relations~\cite{EV1,EV3}.
Next, when the boost is made in the $x$ direction we obtain
\begin{mathletters}
\begin{eqnarray}
{\bf B}^{(3)\,\prime} &=& (-\gamma\beta B^{(0)}, 0, B^{(0)})
\quad ,\quad \mbox{in the coordinates of the old frame} \quad,\\
{\bf B}^{(3)\,\prime} &=& (0, 0, B^{(0)\,\prime})\qquad,\qquad
\mbox{in the coordinates of the new frame} \quad.
\end{eqnarray}
\end{mathletters}
One can see that the transformations (\ref{lt1},\ref{lt2}) are the ones
for a light-like 4-vector of the Minkowski space, formed by $(B^{(0)},\,
{\bf B}^{(3)})$.  They  are similar (while not identical) to the
transformation rules for the spin vector~\cite[Eq.(11.159)]{Jackson}. The
difference with that consideration of a massive particle is caused by
impossibility to find a rest system for the photon which is believed to
move with the invariant velocity $c$. Nevertheless, some relations between
the concept of the Pauli-Lubanski vector of the antisymmetric tensor
fields and the ${\bf B}^{(3)}$ concept  have been found
elsewhere~\cite{EV21,DVOE1,DVOE2}.

I would like to indicate origins of why Dr. Comay achieved the opposite
incorrect result: 1) Obviously, one is not
allowed to identify $B_z$ and ${\bf B}^{(3)}$ (as the authors of
previous papers did, see, e.g., the formula (6) in ref.~\cite{Comay1}),
the first one is an entry
of the antisymmetric tensor and the second one is a 3-vector quantity,
the entry of the 4-vector.  They are different geometric objects. Of
course, the Poynting vector must be perpendicular to the ${\bf E}$ and
${\bf B}$, the Cartesian 3-vectors , whose components are entries of the
antisymmetric tensor field. The ${\bf B}^{(3)}$ vector is a vector of
different nature, while, in its turn, it forms an ``isotopic" vector
with ${\bf B}^{(1)}$ and ${\bf B}^{(2)}$ in a circular {\it complex}
basis\footnote{Let me remind that the Cartesian basis is a pure {\it real}
basis. Introducing {\it complex} vectors we, in fact, enlarge
the space; a number of independent components may increase and, the bases,
in general, are not equivalent mathematically.}
and while its physical effect is similar to that of the Cartesian ${\bf
B}$, namely, magnetization. 2) One is not allowed to forget about the fact
that $B^{(0)}$ is not a {\it scalar quantity}, it is a {\it
zero component} of the 4-vector; so Comay's ``appropriate units"
would transform too from the first to the second
frame.\footnote{Surprisingly, he noted this fact himself in the fourth
Section but ignored it in the Section 3.}
3) As we have found the axial
3-vector ${\bf B}^{(3)}$ is always aligned with the $OZ$ axis in all
frames like the Poynting vector (polar) is, provided that the  ordinary
electric and magnetic fields lie in the $XY$ plane in
these frames.  If this is not the case one can do that by rotation, using
the unitary matrix.   So while the questions raised in the paper
by Comay~\cite{Comay1} may be useful for deeper understanding
of the Evans-Vigier theory and the Relativity Theory
but the conclusion is {\it unreasonable.}
Briefly referring to the paper~[8a] one can
apparently note that, in my opinion, the ${\bf B}^{(3)}$ field is a
property of one photon and when considering the many-photon problem with
various types of polarizations in a superposition the question whether the
circulation of this vector would be different from zero (?) must be
regarded more carefully; furthermore, the applicability of the
dynamical equations to this vector, which Comay refers to, is not obvious
for me.

The conclusion follows in a straightforward manner:
the ${\bf B}^{(3)}$ Evans-Vigier modified
electrodynamics is a relativistic covariant theory if one regards it
mathematically correctly. This construct may be the simplest and
most natural classical representation of a particle spin.
The ${\bf B}-$ cyclic relations manifest relations between Lorentz group
generators answering for the angular momentum~\cite{EV1,EV2,EV21,EV3}.
Moreover, as realized by E. Comay himself ``the modified electrodynamics
relates its longitudinal magnetic field ${\bf B}^{(3)}$ to the
expectation value of the quantum mechanical intrinsic angular momentum
operator". Therefore, recent critics by Profs.  L.  D.  Barron,
A.  D.  Buckingham, E. Comay, D.  Grimes, A.  Lakhtakia
of the Evans-Vigier ${\bf B}^{(3)}$ theory appear to signify
that, in fact, they doubt existence
of the helicity variable for a photon
(additional discrete phase-free variable according to Wigner)
and, hence, all development of physics since its
(helicity) discovery.\footnote{Surprisingly, the opposite
claims (of the {\it pure}
``longitudinal nature" of  the {\it massless} antisymmetric tensor fields)
by several authors are yet another unexplained statements.  That was point
out as long as 1939, by F. Belinfante in the comment to the paper
by Durandin and Erschow [Phys.  Z.  Sowjet.
{\bf 12} (1937) 466]:  ``Three directions
of polarization are possible for a Proca quantum with given momentum and
charge". While the question for neutral particles (self/anti-self charge
conjugate states) should be regarded properly in both the Majorana and the
Dirac constructs, even in this case one can see from the first sight that
those claims of the pure ``longitudinal nature" contradict with a
classical limit and with the Weinberg theorem $B-A=\lambda$, ref.~[10b].
By the way, I do not understand reasons to call this field after the paper
of M.  Kalb and P.  Ramond (1974). As a matter of fact, the antisymmetric
tensor fields (and their ``longitudity") have earlier been investigated by
many authors; first of all by E.  Durandin and A.  Erschow (1937), F.
Belinfante (1939), V.  Ogievetski\u{\i} and I.  Polubarinov (1966), F.
Chang and  F.  G\"ursey (1969), Y. Takahashi and R. Palmer (1970),  K.
Hayashi (1973).} Undoubtedly, such a viewpoint could lead to deep
contradictions with experimental results (the spin-spin interaction, the
inverse Faraday effect, the optical Cotton-Mouton effect, the Tam and
Happer experiment (1977),  etc., etc.).   Whether it
has sufficient reasons?  On an equal footing claims of ``it
is {\it unknowable}" and/or  ``{\it is not fundamental}\," seem to me
to be based on the {\it unknowable} logic.  As opposed to them with
introduction of this variable in a classical
manner~\cite{Belin,Ohan,EV1,EV2,EV21,EV3}
nobody wants to doubt all theoretical results of QED and other gauge
models.  As a matter of fact, existence of the spin variable and of
different polarization states are accounted in calculations of QED matrix
elements.  The proposed development of the Maxwell's theory does not
signify the necessity of rejecting the results which have been obtained in
regions where the old models do work.  Moreover, it was recently
shown~\cite{DVOE1} that both transversal and longitudinal {\it classical}
modes
of electromagnetism are naturally incorporated in the Weinberg
{\it quantum-field} formalism~\cite{Weinberg}.
Thus, the aim of my work (and, I believe, Dr. Evans too) is
to systemize results on the basis of the Poincar\`e group
symmetries, to simplify the theory, to unify
interactions and, perhaps, to predict yet unobserved
phenomena.  One should follow the known advice
of A. Einstein and W. Pauli to build a reliable theory
on the basis of the First Principles,
namely, on the basis of relativistic covariance
(irreducible representations
of the Poincar\'e group) and of causality.
Some progress in this direction has already
been achieved while authors started from different
viewpoints~\cite{DVOE1,DVOE3,EV4}.

I understand that further discussions of the Evans-Vigier model will be
desirable. First of all, the questions arise, whether this theory implies
a photon mass? namely, how does this theory account these effects
(mass appears to manifest itself here in somewhat different form)? if so
what is the massless limit of this theory? what are relations
between the $E(2)$,\, $O(3)$ groups and the group of gauge
transformations of the 4-potential electrodynamics~\cite{Kim}
and of other gauge models?  can a
massless field be particulate?  and, finally, what is the mass itself?  It
is also necessary to give relations of this construct with those presented
by L.  Horwitz, M.  Sachs, A.  Staruszkiewicz, D.  Ahluwalia and myself.
This should be the aim of the forthcoming papers.

I am grateful to Prof. M. Evans for many internet communications on the
concept of the ${\bf B}^{(3)}$ field and estimate his efforts as
considerable (while {\it not always}  agree with him). I acknowledge the
help of Prof.  A.  F. Pashkov, who informed me about the
papers~\cite{Belin,Ohan}, and Prof. D. V. Ahluwalia for his kind comments.

\bigskip

{\it Note Added.} The main addition to the final version of the Comment
by Prof. E. Comay is the claim that the object $({\bf B}^{(0)}, {\bf
B}^{(3)})$ does {\it not} form the (pseudo) 4-vector. I slightly
touched the question of the parity properties in this Note. But, because
after writing this my Comment I has become aware about several more papers
of him, which are aimed at destruction of the Evans-Vigier modified
electrodynamics, and, moreover, the questions of the properties of the
spin-1 massive/massless fields with respect to the discrete symmetry
operations deserve much attention, I think, it would be useful to discuss
the matters of the parity covariance of the ${\bf B}$ Cyclic Relations and
other claims of Prof. E.  Comay in detail in a separate paper.
Here I want to indicate only that in the final version of his Comment (see
this issue of FPL) Professor E.  Comay has made conventions which are
required the rigorous proof:  namely, as a matter of fact he assumed that
the complex-valued ${\bf B}^{(1)}$ and ${\bf B}^{(2)}$ are axial vectors
in the sense that they transform under the space inversion operation as
${\bf B}^{(1)} \rightarrow {\bf B}^{(1)\,\prime}$, and ${\bf B}^{(2)}
\rightarrow {\bf B}^{(2)\,\prime}$ (on the Comay's opinion!).  This wisdom
(if take into account the Lorentz transformation rules of ${\bf B}^{(1)}$
and ${\bf B}^{(2)}$), in fact, puts shades on the rules $\phi_R
\leftrightarrow \phi_L$ with respect to the space inversion
operation~\cite{Ryder} in the $2(2j+1)-$ component theories.

Finally, I want apparently to note that this my paper is NOT a Reply to
the Comay's Comment. In fact, it criticizes both the Evans' work and the
Comay's work. So, it is the Comment and this fact was let to know  to Prof.
Comay before publication (with my permission and along with the draft of
my Comment).  I cannot figure out why does Prof. Comay insist that it is a
Reply?

\end{document}